\documentclass[%
 reprint,
 superscriptaddress,
 longbibliography,
 amsmath,amssymb,
 aps,
 prl,
]{revtex4-1}

\usepackage{amsmath}
\usepackage[colorlinks=true,citecolor=blue]{hyperref}
\usepackage{graphicx}
\usepackage{bm}
\usepackage[dvipsnames]{xcolor}
\usepackage[normalem]{ulem}
\usepackage{gensymb}

\begin{document}

\preprint{APS/123-QED}


\title{Thermal gating of magnon exchange in magnetic multilayers with\\ antiferromagnetic spacers}

\author{D.~M.~Polishchuk}
\email{dpol@kth.se.}
\affiliation{Nanostructure Physics, Royal Institute of Technology, 10691 Stockholm, Sweden}%

\author{Yu.~O.~Tykhonenko-Polishchuk}
\affiliation{Nanostructure Physics, Royal Institute of Technology, 10691 Stockholm, Sweden}%

\author{O.~V.~Gomonay}
\affiliation{Institut f\'ur Physik, Johannes Gutenberg Universit\:at Mainz, D-55099 Mainz, Germany}%

\author{V.~Korenivski}%
\affiliation{Nanostructure Physics, Royal Institute of Technology, 10691 Stockholm, Sweden}%

\date{\today}

\begin{abstract}
We observe a strong thermally-controlled magnon-mediated interlayer coupling of two ferromagnetic layers via an antiferromagnetic spacer in spin-valve type trilayers. The effect manifests itself as a field-induced coherent switching of the two ferromagnets, which can be controlled by varying temperature and the spacer thickness. We explain the observed behavior as due to a strong hybridization of the ferro- and antiferro-magnetic magnon modes in the trilayer at temperatures just below the N\'eel temperature of the antiferromagnetic spacer.
\end{abstract}

\maketitle

\paragraph{Introduction.} Interlayer exchange coupling in magnetic multilayers is one of the key functional properties for spin-electronic applications~\cite{Prinz1998,Gruenberg2001,Zutic2004}. Ferromagnetic interlayer exchange can lead to perpendicular magnetic anisotropy used in ultra-high-density storage media~\cite{Carcia1985,Zeper1989,Sbiaa2011}. Exchange bias found in ferromagnetic/antiferromagnetic (F/AF) bilayers is commonly exploited in spin-valve sensors~\cite{Dieny1994,Monsma1995,Nogues1999}. Indirect interlayer exchange enables the giant magneto-resistive superlattices~\cite{Gruenberg1986,Baibich1988,Binasch1989} and synthetic antiferromagnets~\cite{Jungwirth2016,Duine2018}. Interlayer exchange through antiferromagnetic spacers in F/AF/F trilayers~\cite{Yang2000,Nam2008} makes AF materials attractive as mediators of interlayer coupling in nanostructures~\cite{Kuch2006,Shokr2015,Takei2015,Hagelschuer2016,Wu2018,Tang2019}, which is of special interest for the rapidly developing field of antiferromagnetic spintronics~\cite{Jungwirth2016,Baltz2018} and has the potential to enable ultra-fast (THz), compact, and highly stable to disturbing magnetic fields nanodevices.

Post-fabrication control of the interlayer exchange in a magnetic multilayer can yield additional functionality. For example, thermal control is employed in thermally-assisted magnetic storage media~\cite{Prejbeanu2007} and is also promising for developing spin-thermionic valves and oscillators~\cite{Kadigrobov2010,Kadigrobov2012,Polishchuk2017,Polishchuk2017a,Kravets2017}. Such devices often use the Curie transition in a weakly \emph{ferromagnetic} spacer as the interlayer de/coupling mechanism. An interesting fundamental question is whether an \emph{antiferromagnetic} spacer can be used as the controlling element for on/off switching of interlayer exchange. A successful demonstration could enable a variety of applications in systems where wide-frequency-range oscillators are used, including the presently intensively discussed spin-oscillator-based neuromorphic circuits~\cite{Torrejon2017}. 

\textit{In this Letter}, we demonstrate a method for manipulating de/coupling of thin ferromagnetic layers separated by an antiferromagnetic spacer in F/AF/F trilayers using thermal gating of the magnetic state of the spacer. We use material selective magneto-optical Kerr-effect (MOKE) measurements, enabled by having the F-layers of different magnetic anisotropy (soft F and hard F*), in order to reconstruct the thermo-magnetic state diagram of the trilayers and identify the regions of either coherent or independent field-induced switching of the F-layers. We associate the coherent switching with a strong hybridization of the ferro- and antiferro-magnon modes, which establishes correlations between the ferromagnetic layers. Such correlations are strongly enhanced in the vicinity of the N\'eel transition temperature due to the softening of the antiferromagnetic magnons in this range and can be efficiently switched on and off by varying the sample temperature. The demonstrated \textit{ex-situ} control of interlayer coupling in magnetic nanostructures via thermal gating of antiferromagnetic magnons near room temperature should be promising for spintronic applications.
 
\begin{figure}
\includegraphics[width=6.93 cm]{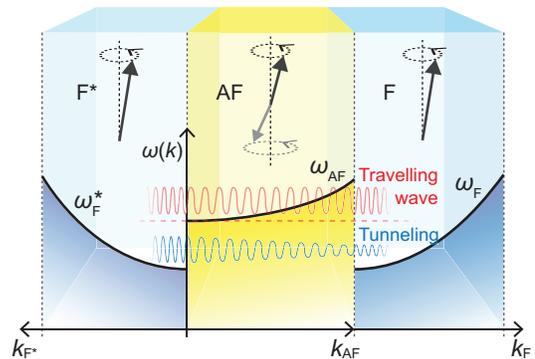}
\caption{Magnon spectra in F*/AF/F (cartoon). F-F* correlations are mediated by two magnon modes that have nonzero amplitude in both F layers. Contribution from traveling modes grows on approach to the N\'eel temperature. Contribution from tunneling modes is higher for thinner spacers.} 

\label{fig_1}
\end{figure}

\paragraph{Theory.} Recently, Cheng \textit{et al.} \cite{Cheng2018b,Cheng2018} developed a model of the magnon-mediated coupling in F/AF/F trilayers by considering excitations in the AF layer only. Here we develop a model that has at its core hybridization of ferro- and antiferro- magnon modes, focusing on the switching conditions for the F and F* layers in F*/AF/F. Assuming the switching is thermally assisted, we search for nonlocalized magnon modes able to establish correlations of the thermal fluctuations in the F and F* layers~\footnote{Even though the structure is metallic, F-F* correlations across the AF spacer due to spin-polarized \emph{conduction electrons} can be neglected since the corresponding spin relaxation length in AF is $\lesssim 1$ nm, much shorter than the AF layer thickness range used in this work; see Merodio \textit{et al.} \href{https://doi.org/10.1063/1.4862971}{Appl. Phys. Lett. \textbf{104}, 032406 (2014)}; Bass \textit{et al.} \href{https://doi.org/10.1016/j.jmmm.2015.12.011}{J. Magn. Magn. Mater. \textbf{408}, 244 (2016)}.}. To calculate the correlation function we consider a model with a collinear uniaxial AF whose magnetic order is characterised  by the N\'eel vector $\mathbf{N}$ and assume parallel alignment of the easy axes in the F and AF layers. We further assume that the exchange coupling at the F*/AF and AF/F interfaces (parametrized by constants $H^*_{b}$ and $H_{b}$, respectively) favours parallel alignment of all the magnetic vectors. Magnons are considered as small excitations $\delta \mathbf{M}^*$, $\delta\mathbf{M}$, and $\delta \mathbf{N}$ over the equilibrium state, with $\mathbf{M}^*\uparrow\uparrow\mathbf{M}\uparrow\uparrow\mathbf{N}$.  We start from the spectra of the decoupled ($H^*_{b}= H_{b}=0$) F*/AF/F layers, which include three magnon branches shown schematically in  Fig.~\ref{fig_1}. Two of them, $\omega^*_\mathrm{F}(k^*_\mathrm{F})$ and $\omega_\mathrm{F}(k_\mathrm{F})$, correspond to the ferromagnetic magnons localised within F* and F layers. Spin polarization of these magnons is directed along $\mathbf{M}^*\uparrow\uparrow\mathbf{M}$. The third branch, $\omega_\mathrm{AF}(k_\mathrm{AF})$, corresponds to the double degenerate modes of the AF magnons with the spin polarization either parallel or antiparalell to $\mathbf{N}$ \cite{Gomonay2018a,Cheng2018}.  The interface exchange coupling, as expected, leads to hybridization of the F and AF magnons with the same spin polarization. The resulting spectra include two types of magnons. The 'tunneling' magnons, whose frequency is lower than $\omega_\mathrm{AF}(0)>\omega_\mathrm{F*}(0),\omega_\mathrm{F}(0)$, are mostly localised within F* or F and decay exponentially into the AF spacer. The corresponding contribution to the correlation function $J\equiv\langle \delta\mathbf{M}^* \delta\mathbf{M} \rangle$ is $\propto H^*_{b}H_{b}\exp(-t/x_\mathrm{DW})$, where $t$ is the thickness of the AF layer and $x_\mathrm{DW}\propto \omega_\mathrm{AF}(0)$ is the characteristic length of the order of the domain wall width. The travelling magnons, whose frequency is higher than $\omega_\mathrm{AF}(0)$, originate from the hybridization of the F and AF magnons in the points of avoided crossing, $\omega_\mathrm{AF}(k_\mathrm{AF})\approx \omega^*_\mathrm{F}(k^*_\mathrm{F}),\omega_\mathrm{F}(k_\mathrm{F})$. Their contribution to the correlation function is suppressed by the thermal distribution factor: $J\propto  H^*_{b}H_{b}\exp\left\{-\hbar\left[\omega_\mathrm{AF}(0)-\omega_\mathrm{F}(0)\right]/k_\mathrm{B}T\right\}$, where $\hbar$ and $k_\mathrm{B}$ are the Planck and the Boltzmann constants, respectively, $T$ is the temperature. Thus, strong correlation between F* and F requires: i) pronounced coupling $H^*_{b},H_{b}$ at both F*/AF and AF/F interfaces; ii) thin AF spacer, $t\ll x_\mathrm{DW}$; and iii) soft AF magnon modes, $\hbar(\omega_\mathrm{AF}(0)-\omega_\mathrm{F}(0))\le k_\mathrm{B}T$. Since, due to the effect of exchange enhancement \cite{Gomonay2018a}, typical AF frequencies are much higher than those in ferromagnets, $\omega_\mathrm{AF}\ge\omega_\mathrm{F}\approx \omega^*_\mathrm{F}$, we can anticipate the correlations to grow mainly in the vicinity of the N\'eel temperature where $\omega_\mathrm{AF}\rightarrow 0$ due to thermally-induced weakening of the long-range AF order. Such softening enhances the thermal population of the traveling magnons and the transmission amplitude of the tunneling magnons, while the interface coupling is large enough to enable hybridization of the F and AF magnons. Far below the N\'eel temperature, the correlation vanishes due to the low population of the travelling magnons caused by increasing $\omega_\mathrm{AF}(0)$, even though the interface coupling is strong. Above the N\'eel temperature, the magnon transport through the spacer as well as the interface coupling are suppressed by the strong spin fluctuations in the paramagnetic spacer and the correlation between the ferromagnetic layers also vanishes.   

\paragraph{Samples and VSM measurements.} We study  F*/AF/F trilayers with the composition of [Fe(6)/Py(3)]/FeMn($t$)/Py(5), where the layer thicknesses in nm are given in parenthesis and FeMn and Py stand for Fe$_{50}$Mn$_{50}$ and Ni$_{81}$Fe$_{19}$ alloys. The Py(5) layer and the [Fe(6)/Py(3)] bilayer are the soft and hard ferromagnetic layers, respectively (hereafter Py and FePy). The Py(3) sub-layer was used to promote the growth of the FeMn layer with the desirable AF properties~\cite{Offi2002,Pan2001} as well as to make the two F/AF interfaces compositionally identical. Since the properties of an antiferromagnet strongly depend on finite-size effects~\cite{Lenz2007}, we fabricated a series of trilayers with different thicknesses of the FeMn spacer ($t =$ 4--15~nm). All the layers were deposited by DC magnetron sputtering (AJA-Orion) at room temperature. In order to induce a preferred magnetization direction, the samples were deposited and subsequently annealed at 250$\degree$C in the presence of a saturating magnetic field.

\begin{figure}
\includegraphics[width=8.5 cm]{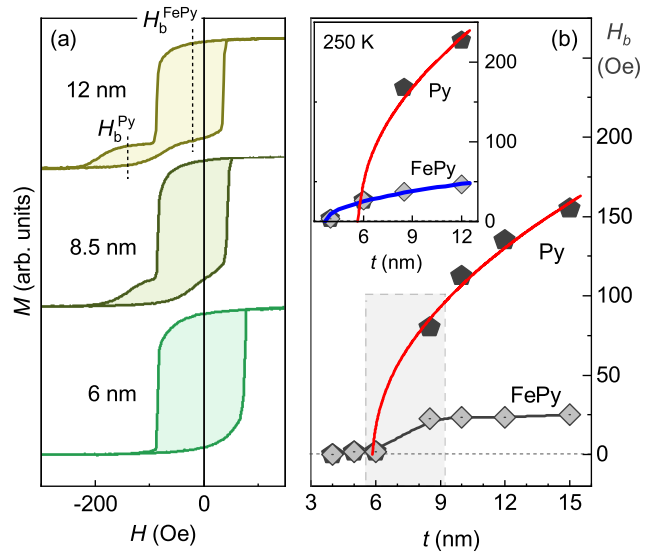}
\caption{Interface exchange coupling in FePy/FeMn($t$)/Py trilayers as a function of FeMn thickness ($t$) at 300~K. (a) VSM magnetization loops for selected AF thicknesses. (b) Thickness dependence of exchange bias field (loop offset) for Py and FePy ($H_b^\mathrm{Py}$ and $H_b^\mathrm{FePy}$) determined as shown in (a); inset shows $H_b$ fields measured at 250~K using MOKE. Solid lines show fitting with power law $H_b\propto\left(t/t_b-1\right)^{1/2}$. Grey bar shows thickness range with pronounced interlayer coupling. }
\label{fig_2}
\end{figure}

Figure~\ref{fig_2}(a) shows typical magnetization loops ($M$-$H$) for the trilayers measured using vibrating sample magnetometry (VSM) at room temperature. For $t >$ 6~nm, the $M$-$H$ loops are composed of two overlapping field-offset loops that correspond to the hard FePy ($H_b^\mathrm{FePy}$) and soft Py ($H_b^\mathrm{Py}$) layers, respectively. The offset vanishes with decreasing $t$, which is due to the known effect of weakening AF order in progressively thinner AF films~\cite{Nogues1999}. A field-centered, single $M$-$H$ loop observed for $t \leq$ 6~nm therefore indicates a vanishing exchange bias and parallel alignment of the outer ferromagnets. Thick AF spacers ($t > $ 10~nm) yield essentially fully decoupled F and F*, where the trilayer $M$-$H$ loop is a superposition of two minor loops with the coercivity and field-offset corresponding to FePy and Py, individually.

We discuss in detail the results for the trilayers with $t =$ 6.0 and 8.5~nm, which are the most representative since their AF-spacers are near the onset of AF-ordering at room temperature, as evidenced by the exchange bias effect; Fig.~\ref{fig_2}(b). The corresponding N\'eel points are conveniently located in the middle of the measurement temperature range, 100--440~K. We note that the results for all studied AF-spacers are qualitatively same, with the thermo-magnetic transition sequentially shifted higher in temperature as the AF thickness is increased.

\begin{figure}
\includegraphics[width=8.5 cm]{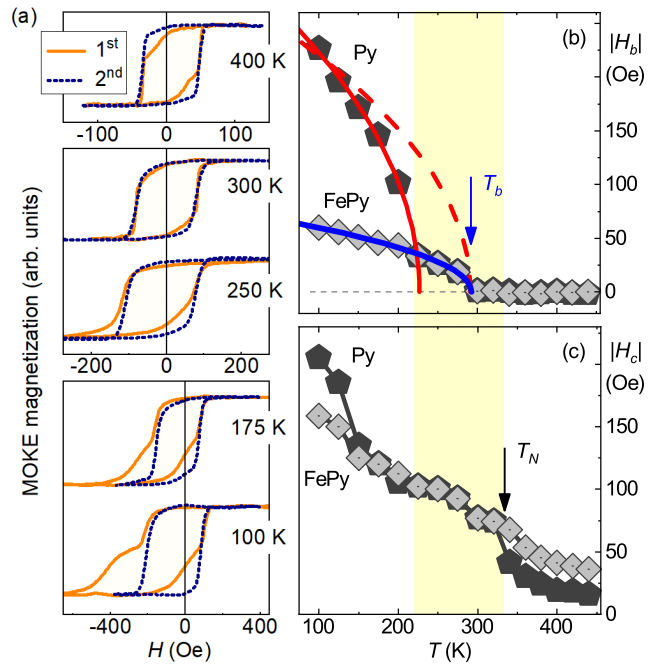}
\caption{Thermo-magnetic transition in FePy/FeMn(6~nm)/ /Py. (a) MOKE loops measured at 1$^\mathrm{st}$ and 2$^\mathrm{nd}$ harmonics of PEM frequency. 2$^\mathrm{nd}$ harmonic is sensitive to FePy layer only, whereas 1$^\mathrm{st}$ harmonic represents integral response of both layers. Temperature dependence of exchange-bias field $H_b$ (b) and coercive field $H_c$ (c) for Py and FePy. Solid lines in (b) are fits using power law $H_b\propto\left(1-T/T_b\right)^{1/2}$; dashed line is behavior expected for standalone Py/FeMn bilayer. Yellow bar shows temperature range of  interlayer coupling.}
\label{fig_3}
\end{figure}

\paragraph{Layer selective MOKE.} We use a photo-elastic modulation (PEM) technique for measuring the longitudinal magneto-optical Kerr-effect (MOKE) of the film samples, which allows us to separate, in a simultaneous measurement, the individual contributions from the soft F and the hard F* layers. The 1$^\mathrm{st}$ harmonic of the PEM-frequency (50~kHz) is proportional to the ellipticity change of the linearly-polarized laser beam, which is sensitive to changes in the total magnetization of a thin-film trilayer. The 2$^\mathrm{nd}$ harmonic (at 100~kHz) is sensitive to the Kerr rotation, dominated in our case by changes in the magnetization of the Fe film. Such material-selectivity allows us to distinguish between the magnetic response of the hard Fe/Py bilayer [2$^\mathrm{nd}$ harmonic; dashed blue line in Fig.~\ref{fig_3}(a)] and the sample's total magnetic response (1$^\mathrm{st}$ harmonic; solid orange line). This MOKE selectivity was calibrated against and confirmed by the VSM measurements on the trilayers (at 300~K), which can distinguish the individual switching of the Py and FePy layers by the respective coercive fields as well as the magnetization step heights (Py and FePy were purposefully made with clearly different magnetic moments, about 1 to 3).

The temperature evolution of the in-plane MOKE $M$-$H$ loops for the FePy/FeMn(6)/Py trilayer ($t =$ 6~nm) is shown in Fig.~\ref{fig_3}(a). At low temperature ($T <$ 220~K), well below the effective $T_\mathrm{N} \approx$ 340~K of the FeMn(6) spacer, the major loop consists of two superposed but clearly distinguishable minor loops, which correspond to Py and FePy. The two minor loops are offset in field due to exchange-pinning by the AF spacer. The non-equal field-offsets indicate that Py and FePy are \emph{pinned individually}, i.e. the \emph{interlayer coupling does not propagate} through the 6-nm-thick AF-spacer at low temperatures where the AF order is strong. With increasing temperature toward $T_\mathrm{N}$, the two minor loops in the combined $M$-$H$ loop (orange, 1$^\mathrm{st}$ harmonic) merge, resulting in a single loop at near room temperature. This indicates that the outer F-layers are \emph{strongly coupled} through the AF-spacer, behaving as a \emph{single ferromagnetic film}. Above $T_\mathrm{N}$, the single loop transforms again into a double-loop [the 1$^\mathrm{st}$-harmonic loop at 400~K in Fig.~\ref{fig_3}(a)], which is now centered at zero field. This double-loop consists of two minor loops of different coercivity. As expected, the two coercive fields are equal to the intrinsic $H_c$ of Py and FePy, which indicates that the ferromagnets are \emph{fully decoupled} and \emph{exchange-unpinned} when the FeMn spacer is paramagnetic.

\paragraph{Magnetic state diagram.} The observed three magnetization regimes in the trilayer represent its thermo-magnetic state diagram and are well illustrated by the temperature dependence of the coercive and exchange-bias fields; Fig.~\ref{fig_3}(b),(c). In particular, the $H_b(T)$ dependence clearly shows the low-$T$ regime with individually exchange-biased Py and FePy and, therefore, negligible ferromagnetic coupling between F and F*. On the other hand, the $H_c(T)$ dependence is very informative in characterizing the high-$T$ regime, above the N\'eel point of the spacer, where Py and FePy switch at their intrinsic coercive fields, again indicating negligible coupling between F and F*. The $H_b$ and $H_c$ values obtained from the 1$^\mathrm{st}$- and 2$^\mathrm{nd}$-harmonic MOKE loops exactly overlap in the intermediate-$T$ range, just below $T_\mathrm{N}$, indicating that the outer ferromagnets are fully coupled in a parallel magnetization state and show coherent switching.

\begin{figure}
\includegraphics[width=8.5 cm]{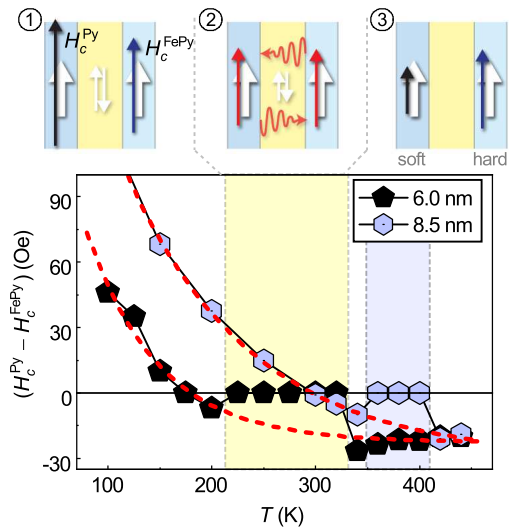}
\caption{Coercive field difference versus temperature for samples with FeMn thickness of 6.0 and 8.5~nm. Dashed lines are 'guides to eye'. Colored bars show regions of interlayer coupling where $\Delta H_c = 0$.
Top panel illustrates three distinct regimes of interaction, with F and F* (i) decoupled and individually exchange-biased ($H_c^\mathrm{Py} \neq H_c^\mathrm{FePy}$, $H_b^\mathrm{Py} \neq H_b^\mathrm{FePy}$, $T \ll T_\mathrm{N}$); (ii) coupled and exchange-pinned as a single layer ($H_c^\mathrm{Py} = H_c^\mathrm{FePy}$, $H_b^\mathrm{Py} = H_b^\mathrm{FePy}$, $T \lesssim T_\mathrm{N}$); and (iii) fully decoupled and unpinned ($H_c^\mathrm{Py} \neq H_c^\mathrm{FePy}$, $H_b^\mathrm{Py,Fe} = 0$, $T > T_\mathrm{N}$).}
\label{fig_4}
\end{figure}

The transition between the three thermo-magnetic regimes is most clearly visualized by the temperature dependence of the coercive field difference $\Delta H_c = (H_c^\mathrm{Py} - H_c^\mathrm{FePy})$, shown in Fig.~\ref{fig_4}. $\Delta H_c$ is non-zero at low-$T$ and high-$T$, which indicates negligible ferromagnetic coupling between F and F*. In the intermediate temperature range, $\Delta H_c$ strictly vanishes (e.g., solid pentagons for the 6~nm sample between 220 and 320~K), indicating ferromagnetically coupled F and F*. 

The coercivity of the Py layer is larger than that of the FePy layer at low temperatures (positive $\Delta H_c$ in Fig.~\ref{fig_4}) since the effect of a AF-surface is known to scale inversely with the F(F*) layer's magnetic moment and the Py layer is both thinner and weaker magnetically. As the temperature is increased toward $T_\mathrm{N}$, the two coercivities merge ($\Delta H_c = 0$) as a result of the AF-mediated ferromagnetic coupling. On further temperature increase, $\Delta H_c$ becomes negative as expected (standalone Py is softer than Fe) and is the same in magnitude for the structures with different AF-spacer thickness (6.0 and 8.5~nm), which additionally indicates that the spacer becomes paramagnetic so F and F* fully decouple. The point of decoupling (loss of AF order or $T_\mathrm{N}$) for the two structures is found to be respectively 340~K and 410~K, which agrees well with the known values for thin-film FeMn~\cite{Lenz2007}. 

Importantly, the magnon-mediated correlations strongly affect the blocking temperature (at which the AF-bias vanishes, $H_b = 0$) of the F and F* layers. We determine the blocking temperature $T_b$ by fitting $H_b(T)$ to a power law consistent with the Landau phase transition theory [$\propto\left(1-T/T_b\right)^{1/2}$; see Fig.~\ref{fig_3}(b)]. While the individual bilayers Py/FeMn and FePy/FeMn have identical interfaces and thus should have the same $T_b$ [vertical arrow in Fig.~\ref{fig_3}(b)], the trilayer data show a large drop in the effective blocking temperature of the soft layer, $T^\mathrm{soft}_b <T_b$. This drop (from about 300 to 220~K) is a direct effect of the enhanced fluctuations in the soft layer due to the nonlocalized magnon modes excited in the hard layer. In contrast, and in line with expectations, the hard layer is much less sensitive to the influence of the soft-layer's magnons (its effective $T^\mathrm{hard}_b\approx T_b$) and, therefore, defines the blocking temperature of the trilayer. As  $T^\mathrm{hard}_b$ is found midway the region of strong interlayer correlations (bound by $T^\mathrm{soft}_b$ and $T_N$) the coherent switching is, in fact, found to take place with ($T^\mathrm{soft}_b <T<T^\mathrm{hard}_b$) or without ($T^\mathrm{hard}_b <T<T_N$) exchange bias. Remarkably, the temperature and AF-thickness dependence of the observed interlayer coupling are complementary, which becomes evident upon comparing Fig.~\ref{fig_3}(b) with Fig.~\ref{fig_2}(b), where $H_b(t)$ and $H_b(T)$ show the same power law dependence.

\paragraph{Conclusion.} The fact that F*/AF/F trilayers are ferromagnetically coupled at temperatures just below $T_\mathrm{N}$ of the AF spacer indicates a thermo-magnetic origin of the effect. At $T \lesssim T_\mathrm{N}$, the antiferromaget experiences strong thermal agitation resulting in a softening of its magnon modes toward
the low-GHz range, which is characteristic of magnons in ferromagnets. The low-frequency magnons can then propagate through the spacer, exchanging spin between the outer ferromagnets. At significantly lower temperatures ($T \ll T_\mathrm{N}$), 
the magnon gap in the AF-spacer broadens toward the THz range and
acts as a potential barrier to interlayer magnon exchange. At high temperatures ($T > T_\mathrm{N}$), the spacer is fully spin disordered and  unable to transmit magnons. Thus, from the viewpoint of correlations between the outer ferromagnets, the AF spacer undergoes a thermal transition of type insulator-conductor-insulator, reflected in the three distinct thermo-magnetic regimes found in the system. 


\begin{acknowledgments}
Support from the Swedish Research Council (VR:2018-03526) and Olle Engkvist Foundation (SOEB:207-0460) are gratefully acknowledged. O.G. acknowledges support from the Alexander von Humboldt Foundation, the ERC Synergy Grant SC2 (No. 610115), and the Deutsche Forschungsgemeinschaft (DFG, German Research Foundation) - TRR 173 – 268565370 (project B12).
\end{acknowledgments}

\bibliography{refMagnEx}

\begin{thebibliography}{37}%
\makeatletter
\providecommand \@ifxundefined [1]{%
 \@ifx{#1\undefined}
}%
\providecommand \@ifnum [1]{%
 \ifnum #1\expandafter \@firstoftwo
 \else \expandafter \@secondoftwo
 \fi
}%
\providecommand \@ifx [1]{%
 \ifx #1\expandafter \@firstoftwo
 \else \expandafter \@secondoftwo
 \fi
}%
\providecommand \natexlab [1]{#1}%
\providecommand \enquote  [1]{``#1''}%
\providecommand \bibnamefont  [1]{#1}%
\providecommand \bibfnamefont [1]{#1}%
\providecommand \citenamefont [1]{#1}%
\providecommand \href@noop [0]{\@secondoftwo}%
\providecommand \href [0]{\begingroup \@sanitize@url \@href}%
\providecommand \@href[1]{\@@startlink{#1}\@@href}%
\providecommand \@@href[1]{\endgroup#1\@@endlink}%
\providecommand \@sanitize@url [0]{\catcode `\\12\catcode `\$12\catcode
  `\&12\catcode `\#12\catcode `\^12\catcode `\_12\catcode `\%12\relax}%
\providecommand \@@startlink[1]{}%
\providecommand \@@endlink[0]{}%
\providecommand \url  [0]{\begingroup\@sanitize@url \@url }%
\providecommand \@url [1]{\endgroup\@href {#1}{\urlprefix }}%
\providecommand \urlprefix  [0]{URL }%
\providecommand \Eprint [0]{\href }%
\providecommand \doibase [0]{http://dx.doi.org/}%
\providecommand \selectlanguage [0]{\@gobble}%
\providecommand \bibinfo  [0]{\@secondoftwo}%
\providecommand \bibfield  [0]{\@secondoftwo}%
\providecommand \translation [1]{[#1]}%
\providecommand \BibitemOpen [0]{}%
\providecommand \bibitemStop [0]{}%
\providecommand \bibitemNoStop [0]{.\EOS\space}%
\providecommand \EOS [0]{\spacefactor3000\relax}%
\providecommand \BibitemShut  [1]{\csname bibitem#1\endcsname}%
\let\auto@bib@innerbib\@empty
\bibitem [{\citenamefont {Prinz}(1998)}]{Prinz1998}%
  \BibitemOpen
  \bibfield  {author} {\bibinfo {author} {\bibfnamefont {G.~A.}\ \bibnamefont
  {Prinz}},\ }\bibfield  {title} {\enquote {\bibinfo {title}
  {Magnetoelectronics},}\ }\href {\doibase 10.1126/science.282.5394.1660}
  {\bibfield  {journal} {\bibinfo  {journal} {Science}\ }\textbf {\bibinfo
  {volume} {282}},\ \bibinfo {pages} {1660--1663} (\bibinfo {year}
  {1998})}\BibitemShut {NoStop}%
\bibitem [{\citenamefont {Gr\"unberg}(2001)}]{Gruenberg2001}%
  \BibitemOpen
  \bibfield  {author} {\bibinfo {author} {\bibfnamefont {Peter}\ \bibnamefont
  {Gr\"unberg}},\ }\bibfield  {title} {\enquote {\bibinfo {title} {Layered
  magnetic structures: History, highlights, applications},}\ }\href {\doibase
  10.1063/1.1381100} {\bibfield  {journal} {\bibinfo  {journal} {Physics
  Today}\ }\textbf {\bibinfo {volume} {54}},\ \bibinfo {pages} {31--37}
  (\bibinfo {year} {2001})}\BibitemShut {NoStop}%
\bibitem [{\citenamefont {{\v{Z}}uti{\'{c}}}\ \emph {et~al.}(2004)\citenamefont
  {{\v{Z}}uti{\'{c}}}, \citenamefont {Fabian},\ and\ \citenamefont
  {Sarma}}]{Zutic2004}%
  \BibitemOpen
  \bibfield  {author} {\bibinfo {author} {\bibfnamefont {Igor}\ \bibnamefont
  {{\v{Z}}uti{\'{c}}}}, \bibinfo {author} {\bibfnamefont {Jaroslav}\
  \bibnamefont {Fabian}}, \ and\ \bibinfo {author} {\bibfnamefont {S.~Das}\
  \bibnamefont {Sarma}},\ }\bibfield  {title} {\enquote {\bibinfo {title}
  {Spintronics: Fundamentals and applications},}\ }\href {\doibase
  10.1103/revmodphys.76.323} {\bibfield  {journal} {\bibinfo  {journal}
  {Reviews of Modern Physics}\ }\textbf {\bibinfo {volume} {76}},\ \bibinfo
  {pages} {323--410} (\bibinfo {year} {2004})}\BibitemShut {NoStop}%
\bibitem [{\citenamefont {Carcia}\ \emph {et~al.}(1985)\citenamefont {Carcia},
  \citenamefont {Meinhaldt},\ and\ \citenamefont {Suna}}]{Carcia1985}%
  \BibitemOpen
  \bibfield  {author} {\bibinfo {author} {\bibfnamefont {P.~F.}\ \bibnamefont
  {Carcia}}, \bibinfo {author} {\bibfnamefont {A.~D.}\ \bibnamefont
  {Meinhaldt}}, \ and\ \bibinfo {author} {\bibfnamefont {A.}~\bibnamefont
  {Suna}},\ }\bibfield  {title} {\enquote {\bibinfo {title} {Perpendicular
  magnetic anisotropy in pd/co thin film layered structures},}\ }\href
  {\doibase 10.1063/1.96254} {\bibfield  {journal} {\bibinfo  {journal}
  {Applied Physics Letters}\ }\textbf {\bibinfo {volume} {47}},\ \bibinfo
  {pages} {178--180} (\bibinfo {year} {1985})}\BibitemShut {NoStop}%
\bibitem [{\citenamefont {Zeper}\ \emph {et~al.}(1989)\citenamefont {Zeper},
  \citenamefont {Greidanus}, \citenamefont {Carcia},\ and\ \citenamefont
  {Fincher}}]{Zeper1989}%
  \BibitemOpen
  \bibfield  {author} {\bibinfo {author} {\bibfnamefont {W.~B.}\ \bibnamefont
  {Zeper}}, \bibinfo {author} {\bibfnamefont {F.~J. A.~M.}\ \bibnamefont
  {Greidanus}}, \bibinfo {author} {\bibfnamefont {P.~F.}\ \bibnamefont
  {Carcia}}, \ and\ \bibinfo {author} {\bibfnamefont {C.~R.}\ \bibnamefont
  {Fincher}},\ }\bibfield  {title} {\enquote {\bibinfo {title} {Perpendicular
  magnetic anisotropy and magneto-optical kerr effect of vapor-deposited
  co/pt-layered structures},}\ }\href {\doibase 10.1063/1.343189} {\bibfield
  {journal} {\bibinfo  {journal} {Journal of Applied Physics}\ }\textbf
  {\bibinfo {volume} {65}},\ \bibinfo {pages} {4971--4975} (\bibinfo {year}
  {1989})}\BibitemShut {NoStop}%
\bibitem [{\citenamefont {Sbiaa}\ \emph {et~al.}(2011)\citenamefont {Sbiaa},
  \citenamefont {Meng},\ and\ \citenamefont {Piramanayagam}}]{Sbiaa2011}%
  \BibitemOpen
  \bibfield  {author} {\bibinfo {author} {\bibfnamefont {R.}~\bibnamefont
  {Sbiaa}}, \bibinfo {author} {\bibfnamefont {H.}~\bibnamefont {Meng}}, \ and\
  \bibinfo {author} {\bibfnamefont {S.~N.}\ \bibnamefont {Piramanayagam}},\
  }\bibfield  {title} {\enquote {\bibinfo {title} {Materials with perpendicular
  magnetic anisotropy for magnetic random access memory},}\ }\href {\doibase
  10.1002/pssr.201105420} {\bibfield  {journal} {\bibinfo  {journal} {Physica
  Status Solidi ({RRL}) - Rapid Research Letters}\ }\textbf {\bibinfo {volume}
  {5}},\ \bibinfo {pages} {413--419} (\bibinfo {year} {2011})}\BibitemShut
  {NoStop}%
\bibitem [{\citenamefont {Dieny}(1994)}]{Dieny1994}%
  \BibitemOpen
  \bibfield  {author} {\bibinfo {author} {\bibfnamefont {B.}~\bibnamefont
  {Dieny}},\ }\bibfield  {title} {\enquote {\bibinfo {title} {Giant
  magnetoresistance in spin-valve multilayers},}\ }\href {\doibase
  10.1016/0304-8853(94)00356-4} {\bibfield  {journal} {\bibinfo  {journal}
  {Journal of Magnetism and Magnetic Materials}\ }\textbf {\bibinfo {volume}
  {136}},\ \bibinfo {pages} {335--359} (\bibinfo {year} {1994})}\BibitemShut
  {NoStop}%
\bibitem [{\citenamefont {Monsma}\ \emph {et~al.}(1995)\citenamefont {Monsma},
  \citenamefont {Lodder}, \citenamefont {Popma},\ and\ \citenamefont
  {Dieny}}]{Monsma1995}%
  \BibitemOpen
  \bibfield  {author} {\bibinfo {author} {\bibfnamefont {D.~J.}\ \bibnamefont
  {Monsma}}, \bibinfo {author} {\bibfnamefont {J.~C.}\ \bibnamefont {Lodder}},
  \bibinfo {author} {\bibfnamefont {Th. J.~A.}\ \bibnamefont {Popma}}, \ and\
  \bibinfo {author} {\bibfnamefont {B.}~\bibnamefont {Dieny}},\ }\bibfield
  {title} {\enquote {\bibinfo {title} {Perpendicular hot electron spin-valve
  effect in a new magnetic field sensor: The spin-valve transistor},}\ }\href
  {\doibase 10.1103/physrevlett.74.5260} {\bibfield  {journal} {\bibinfo
  {journal} {Physical Review Letters}\ }\textbf {\bibinfo {volume} {74}},\
  \bibinfo {pages} {5260--5263} (\bibinfo {year} {1995})}\BibitemShut {NoStop}%
\bibitem [{\citenamefont {Nogu{\'{e}}s}\ and\ \citenamefont
  {Schuller}(1999)}]{Nogues1999}%
  \BibitemOpen
  \bibfield  {author} {\bibinfo {author} {\bibfnamefont {J}~\bibnamefont
  {Nogu{\'{e}}s}}\ and\ \bibinfo {author} {\bibfnamefont {Ivan~K}\ \bibnamefont
  {Schuller}},\ }\bibfield  {title} {\enquote {\bibinfo {title} {Exchange
  bias},}\ }\href {\doibase 10.1016/s0304-8853(98)00266-2} {\bibfield
  {journal} {\bibinfo  {journal} {Journal of Magnetism and Magnetic Materials}\
  }\textbf {\bibinfo {volume} {192}},\ \bibinfo {pages} {203--232} (\bibinfo
  {year} {1999})}\BibitemShut {NoStop}%
\bibitem [{\citenamefont {Gr\"unberg}\ \emph {et~al.}(1986)\citenamefont
  {Gr\"unberg}, \citenamefont {Schreiber}, \citenamefont {Pang}, \citenamefont
  {Brodsky},\ and\ \citenamefont {Sowers}}]{Gruenberg1986}%
  \BibitemOpen
  \bibfield  {author} {\bibinfo {author} {\bibfnamefont {P.}~\bibnamefont
  {Gr\"unberg}}, \bibinfo {author} {\bibfnamefont {R.}~\bibnamefont
  {Schreiber}}, \bibinfo {author} {\bibfnamefont {Y.}~\bibnamefont {Pang}},
  \bibinfo {author} {\bibfnamefont {M.~B.}\ \bibnamefont {Brodsky}}, \ and\
  \bibinfo {author} {\bibfnamefont {H.}~\bibnamefont {Sowers}},\ }\bibfield
  {title} {\enquote {\bibinfo {title} {Layered magnetic structures: Evidence
  for antiferromagnetic coupling of {Fe} layers across {Cr} interlayers},}\
  }\href {\doibase 10.1103/physrevlett.57.2442} {\bibfield  {journal} {\bibinfo
   {journal} {Physical Review Letters}\ }\textbf {\bibinfo {volume} {57}},\
  \bibinfo {pages} {2442--2445} (\bibinfo {year} {1986})}\BibitemShut {NoStop}%
\bibitem [{\citenamefont {Baibich}\ \emph {et~al.}(1988)\citenamefont
  {Baibich}, \citenamefont {Broto}, \citenamefont {Fert}, \citenamefont {Dau},
  \citenamefont {Petroff}, \citenamefont {Etienne}, \citenamefont {Creuzet},
  \citenamefont {Friederich},\ and\ \citenamefont {Chazelas}}]{Baibich1988}%
  \BibitemOpen
  \bibfield  {author} {\bibinfo {author} {\bibfnamefont {M.~N.}\ \bibnamefont
  {Baibich}}, \bibinfo {author} {\bibfnamefont {J.~M.}\ \bibnamefont {Broto}},
  \bibinfo {author} {\bibfnamefont {A.}~\bibnamefont {Fert}}, \bibinfo {author}
  {\bibfnamefont {F.~Nguyen~Van}\ \bibnamefont {Dau}}, \bibinfo {author}
  {\bibfnamefont {F.}~\bibnamefont {Petroff}}, \bibinfo {author} {\bibfnamefont
  {P.}~\bibnamefont {Etienne}}, \bibinfo {author} {\bibfnamefont
  {G.}~\bibnamefont {Creuzet}}, \bibinfo {author} {\bibfnamefont
  {A.}~\bibnamefont {Friederich}}, \ and\ \bibinfo {author} {\bibfnamefont
  {J.}~\bibnamefont {Chazelas}},\ }\bibfield  {title} {\enquote {\bibinfo
  {title} {Giant magnetoresistance of (001){Fe}/(001){Cr} magnetic
  superlattices},}\ }\href {\doibase 10.1103/physrevlett.61.2472} {\bibfield
  {journal} {\bibinfo  {journal} {Physical Review Letters}\ }\textbf {\bibinfo
  {volume} {61}},\ \bibinfo {pages} {2472--2475} (\bibinfo {year}
  {1988})}\BibitemShut {NoStop}%
\bibitem [{\citenamefont {Binasch}\ \emph {et~al.}(1989)\citenamefont
  {Binasch}, \citenamefont {Grünberg}, \citenamefont {Saurenbach},\ and\
  \citenamefont {Zinn}}]{Binasch1989}%
  \BibitemOpen
  \bibfield  {author} {\bibinfo {author} {\bibfnamefont {G.}~\bibnamefont
  {Binasch}}, \bibinfo {author} {\bibfnamefont {P.}~\bibnamefont {Grünberg}},
  \bibinfo {author} {\bibfnamefont {F.}~\bibnamefont {Saurenbach}}, \ and\
  \bibinfo {author} {\bibfnamefont {W.}~\bibnamefont {Zinn}},\ }\bibfield
  {title} {\enquote {\bibinfo {title} {Enhanced magnetoresistance in layered
  magnetic structures with antiferromagnetic interlayer exchange},}\ }\href
  {\doibase 10.1103/physrevb.39.4828} {\bibfield  {journal} {\bibinfo
  {journal} {Physical Review B}\ }\textbf {\bibinfo {volume} {39}},\ \bibinfo
  {pages} {4828--4830} (\bibinfo {year} {1989})}\BibitemShut {NoStop}%
\bibitem [{\citenamefont {Jungwirth}\ \emph {et~al.}(2016)\citenamefont
  {Jungwirth}, \citenamefont {Marti}, \citenamefont {Wadley},\ and\
  \citenamefont {Wunderlich}}]{Jungwirth2016}%
  \BibitemOpen
  \bibfield  {author} {\bibinfo {author} {\bibfnamefont {T.}~\bibnamefont
  {Jungwirth}}, \bibinfo {author} {\bibfnamefont {X.}~\bibnamefont {Marti}},
  \bibinfo {author} {\bibfnamefont {P.}~\bibnamefont {Wadley}}, \ and\ \bibinfo
  {author} {\bibfnamefont {J.}~\bibnamefont {Wunderlich}},\ }\bibfield  {title}
  {\enquote {\bibinfo {title} {Antiferromagnetic spintronics},}\ }\href
  {\doibase 10.1038/nnano.2016.18} {\bibfield  {journal} {\bibinfo  {journal}
  {Nature Nanotechnology}\ }\textbf {\bibinfo {volume} {11}},\ \bibinfo {pages}
  {231--241} (\bibinfo {year} {2016})}\BibitemShut {NoStop}%
\bibitem [{\citenamefont {Duine}\ \emph {et~al.}(2018)\citenamefont {Duine},
  \citenamefont {Lee}, \citenamefont {Parkin},\ and\ \citenamefont
  {Stiles}}]{Duine2018}%
  \BibitemOpen
  \bibfield  {author} {\bibinfo {author} {\bibfnamefont {R.~A.}\ \bibnamefont
  {Duine}}, \bibinfo {author} {\bibfnamefont {Kyung-Jin}\ \bibnamefont {Lee}},
  \bibinfo {author} {\bibfnamefont {Stuart S.~P.}\ \bibnamefont {Parkin}}, \
  and\ \bibinfo {author} {\bibfnamefont {M.~D.}\ \bibnamefont {Stiles}},\
  }\bibfield  {title} {\enquote {\bibinfo {title} {Synthetic antiferromagnetic
  spintronics},}\ }\href {\doibase 10.1038/s41567-018-0050-y} {\bibfield
  {journal} {\bibinfo  {journal} {Nature Physics}\ }\textbf {\bibinfo {volume}
  {14}},\ \bibinfo {pages} {217--219} (\bibinfo {year} {2018})}\BibitemShut
  {NoStop}%
\bibitem [{\citenamefont {Yang}\ and\ \citenamefont {Chien}(2000)}]{Yang2000}%
  \BibitemOpen
  \bibfield  {author} {\bibinfo {author} {\bibfnamefont {F.~Y.}\ \bibnamefont
  {Yang}}\ and\ \bibinfo {author} {\bibfnamefont {C.~L.}\ \bibnamefont
  {Chien}},\ }\bibfield  {title} {\enquote {\bibinfo {title} {Spiraling spin
  structure in an exchange-coupled antiferromagnetic layer},}\ }\href {\doibase
  10.1103/physrevlett.85.2597} {\bibfield  {journal} {\bibinfo  {journal}
  {Physical Review Letters}\ }\textbf {\bibinfo {volume} {85}},\ \bibinfo
  {pages} {2597--2600} (\bibinfo {year} {2000})}\BibitemShut {NoStop}%
\bibitem [{\citenamefont {Nam}\ \emph {et~al.}(2008)\citenamefont {Nam},
  \citenamefont {Chen}, \citenamefont {West}, \citenamefont {Kirkwood},
  \citenamefont {Lu},\ and\ \citenamefont {Wolf}}]{Nam2008}%
  \BibitemOpen
  \bibfield  {author} {\bibinfo {author} {\bibfnamefont {D.~N.~H.}\
  \bibnamefont {Nam}}, \bibinfo {author} {\bibfnamefont {W.}~\bibnamefont
  {Chen}}, \bibinfo {author} {\bibfnamefont {K.~G.}\ \bibnamefont {West}},
  \bibinfo {author} {\bibfnamefont {D.~M.}\ \bibnamefont {Kirkwood}}, \bibinfo
  {author} {\bibfnamefont {J.}~\bibnamefont {Lu}}, \ and\ \bibinfo {author}
  {\bibfnamefont {S.~A.}\ \bibnamefont {Wolf}},\ }\bibfield  {title} {\enquote
  {\bibinfo {title} {Propagation of exchange bias in {CoFe}/{FeMn}/{CoFe}
  trilayers},}\ }\href {\doibase 10.1063/1.2999626} {\bibfield  {journal}
  {\bibinfo  {journal} {Applied Physics Letters}\ }\textbf {\bibinfo {volume}
  {93}},\ \bibinfo {pages} {152504} (\bibinfo {year} {2008})}\BibitemShut
  {NoStop}%
\bibitem [{\citenamefont {Kuch}\ \emph {et~al.}(2006)\citenamefont {Kuch},
  \citenamefont {Chelaru}, \citenamefont {Offi}, \citenamefont {Wang},
  \citenamefont {Kotsugi},\ and\ \citenamefont {Kirschner}}]{Kuch2006}%
  \BibitemOpen
  \bibfield  {author} {\bibinfo {author} {\bibfnamefont {W.}~\bibnamefont
  {Kuch}}, \bibinfo {author} {\bibfnamefont {L.~I.}\ \bibnamefont {Chelaru}},
  \bibinfo {author} {\bibfnamefont {F.}~\bibnamefont {Offi}}, \bibinfo {author}
  {\bibfnamefont {J.}~\bibnamefont {Wang}}, \bibinfo {author} {\bibfnamefont
  {M.}~\bibnamefont {Kotsugi}}, \ and\ \bibinfo {author} {\bibfnamefont
  {J.}~\bibnamefont {Kirschner}},\ }\bibfield  {title} {\enquote {\bibinfo
  {title} {Tuning the magnetic coupling across ultrathin antiferromagnetic
  films by controlling atomic-scale roughness},}\ }\href {\doibase
  10.1038/nmat1548} {\bibfield  {journal} {\bibinfo  {journal} {Nature
  Materials}\ }\textbf {\bibinfo {volume} {5}},\ \bibinfo {pages} {128--133}
  (\bibinfo {year} {2006})}\BibitemShut {NoStop}%
\bibitem [{\citenamefont {Shokr}\ \emph {et~al.}(2015)\citenamefont {Shokr},
  \citenamefont {Erkovan}, \citenamefont {Sandig},\ and\ \citenamefont
  {Kuch}}]{Shokr2015}%
  \BibitemOpen
  \bibfield  {author} {\bibinfo {author} {\bibfnamefont {Y.~A.}\ \bibnamefont
  {Shokr}}, \bibinfo {author} {\bibfnamefont {M.}~\bibnamefont {Erkovan}},
  \bibinfo {author} {\bibfnamefont {O.}~\bibnamefont {Sandig}}, \ and\ \bibinfo
  {author} {\bibfnamefont {W.}~\bibnamefont {Kuch}},\ }\bibfield  {title}
  {\enquote {\bibinfo {title} {Temperature-induced sign change of the magnetic
  interlayer coupling in {Ni}/{Ni}$_{25}${Mn}$_{75}$/{Ni} trilayers on
  {Cu}$_3${Au}(001)},}\ }\href {\doibase 10.1063/1.4919597} {\bibfield
  {journal} {\bibinfo  {journal} {Journal of Applied Physics}\ }\textbf
  {\bibinfo {volume} {117}},\ \bibinfo {pages} {175302} (\bibinfo {year}
  {2015})}\BibitemShut {NoStop}%
\bibitem [{\citenamefont {Takei}\ \emph {et~al.}(2015)\citenamefont {Takei},
  \citenamefont {Moriyama}, \citenamefont {Ono},\ and\ \citenamefont
  {Tserkovnyak}}]{Takei2015}%
  \BibitemOpen
  \bibfield  {author} {\bibinfo {author} {\bibfnamefont {S.}~\bibnamefont
  {Takei}}, \bibinfo {author} {\bibfnamefont {T.}~\bibnamefont {Moriyama}},
  \bibinfo {author} {\bibfnamefont {T.}~\bibnamefont {Ono}}, \ and\ \bibinfo
  {author} {\bibfnamefont {Ya.}\ \bibnamefont {Tserkovnyak}},\ }\bibfield
  {title} {\enquote {\bibinfo {title} {Antiferromagnet-mediated spin transfer
  between a metal and a ferromagnet},}\ }\href {\doibase
  10.1103/physrevb.92.020409} {\bibfield  {journal} {\bibinfo  {journal}
  {Physical Review B}\ }\textbf {\bibinfo {volume} {92}},\ \bibinfo {pages}
  {020409(R)} (\bibinfo {year} {2015})}\BibitemShut {NoStop}%
\bibitem [{\citenamefont {Hagelschuer}\ \emph {et~al.}(2016)\citenamefont
  {Hagelschuer}, \citenamefont {Shokr},\ and\ \citenamefont
  {Kuch}}]{Hagelschuer2016}%
  \BibitemOpen
  \bibfield  {author} {\bibinfo {author} {\bibfnamefont {T.}~\bibnamefont
  {Hagelschuer}}, \bibinfo {author} {\bibfnamefont {Y.~A.}\ \bibnamefont
  {Shokr}}, \ and\ \bibinfo {author} {\bibfnamefont {W.}~\bibnamefont {Kuch}},\
  }\bibfield  {title} {\enquote {\bibinfo {title} {Spin-state transition in
  antiferromagnetic {Ni}$_{0.4}${Mn}$_{0.6}$ films in {Ni}/{NiMn}/{Ni}
  trilayers on {Cu}(001)},}\ }\href {\doibase 10.1103/physrevb.93.054428}
  {\bibfield  {journal} {\bibinfo  {journal} {Physical Review B}\ }\textbf
  {\bibinfo {volume} {93}},\ \bibinfo {pages} {054428} (\bibinfo {year}
  {2016})}\BibitemShut {NoStop}%
\bibitem [{\citenamefont {Wu}\ \emph {et~al.}(2018)\citenamefont {Wu},
  \citenamefont {Huang}, \citenamefont {Fang}, \citenamefont {Yang},
  \citenamefont {Wan}, \citenamefont {Yu}, \citenamefont {Feng}, \citenamefont
  {Wei},\ and\ \citenamefont {Han}}]{Wu2018}%
  \BibitemOpen
  \bibfield  {author} {\bibinfo {author} {\bibfnamefont {H.}~\bibnamefont
  {Wu}}, \bibinfo {author} {\bibfnamefont {L.}~\bibnamefont {Huang}}, \bibinfo
  {author} {\bibfnamefont {C.}~\bibnamefont {Fang}}, \bibinfo {author}
  {\bibfnamefont {B.{\hspace{0.167em}}S.}\ \bibnamefont {Yang}}, \bibinfo
  {author} {\bibfnamefont {C.{\hspace{0.167em}}H.}\ \bibnamefont {Wan}},
  \bibinfo {author} {\bibfnamefont {G.{\hspace{0.167em}}Q.}\ \bibnamefont
  {Yu}}, \bibinfo {author} {\bibfnamefont {J.{\hspace{0.167em}}F.}\
  \bibnamefont {Feng}}, \bibinfo {author} {\bibfnamefont
  {H.{\hspace{0.167em}}X.}\ \bibnamefont {Wei}}, \ and\ \bibinfo {author}
  {\bibfnamefont {X.{\hspace{0.167em}}F.}\ \bibnamefont {Han}},\ }\bibfield
  {title} {\enquote {\bibinfo {title} {Magnon valve effect between two magnetic
  insulators},}\ }\href {\doibase 10.1103/physrevlett.120.097205} {\bibfield
  {journal} {\bibinfo  {journal} {Physical Review Letters}\ }\textbf {\bibinfo
  {volume} {120}},\ \bibinfo {pages} {097205} (\bibinfo {year}
  {2018})}\BibitemShut {NoStop}%
\bibitem [{\citenamefont {Tang}\ and\ \citenamefont {Han}(2019)}]{Tang2019}%
  \BibitemOpen
  \bibfield  {author} {\bibinfo {author} {\bibfnamefont {Ping}\ \bibnamefont
  {Tang}}\ and\ \bibinfo {author} {\bibfnamefont {X.~F.}\ \bibnamefont {Han}},\
  }\bibfield  {title} {\enquote {\bibinfo {title} {Magnon resonant tunneling
  effect in double-barrier insulating magnon junctions and magnon field effect
  transistor},}\ }\href {\doibase 10.1103/physrevb.99.054401} {\bibfield
  {journal} {\bibinfo  {journal} {Physical Review B}\ }\textbf {\bibinfo
  {volume} {99}},\ \bibinfo {pages} {054401} (\bibinfo {year}
  {2019})}\BibitemShut {NoStop}%
\bibitem [{\citenamefont {Baltz}\ \emph {et~al.}(2018)\citenamefont {Baltz},
  \citenamefont {Manchon}, \citenamefont {Tsoi}, \citenamefont {Moriyama},
  \citenamefont {Ono},\ and\ \citenamefont {Tserkovnyak}}]{Baltz2018}%
  \BibitemOpen
  \bibfield  {author} {\bibinfo {author} {\bibfnamefont {V.}~\bibnamefont
  {Baltz}}, \bibinfo {author} {\bibfnamefont {A.}~\bibnamefont {Manchon}},
  \bibinfo {author} {\bibfnamefont {M.}~\bibnamefont {Tsoi}}, \bibinfo {author}
  {\bibfnamefont {T.}~\bibnamefont {Moriyama}}, \bibinfo {author}
  {\bibfnamefont {T.}~\bibnamefont {Ono}}, \ and\ \bibinfo {author}
  {\bibfnamefont {Y.}~\bibnamefont {Tserkovnyak}},\ }\bibfield  {title}
  {\enquote {\bibinfo {title} {Antiferromagnetic spintronics},}\ }\href
  {\doibase 10.1103/revmodphys.90.015005} {\bibfield  {journal} {\bibinfo
  {journal} {Reviews of Modern Physics}\ }\textbf {\bibinfo {volume} {90}},\
  \bibinfo {pages} {015005} (\bibinfo {year} {2018})}\BibitemShut {NoStop}%
\bibitem [{\citenamefont {Prejbeanu}\ \emph {et~al.}(2007)\citenamefont
  {Prejbeanu}, \citenamefont {Kerekes}, \citenamefont {Sousa}, \citenamefont
  {Sibuet}, \citenamefont {Redon}, \citenamefont {Dieny},\ and\ \citenamefont
  {Nozi{\`{e}}res}}]{Prejbeanu2007}%
  \BibitemOpen
  \bibfield  {author} {\bibinfo {author} {\bibfnamefont {I~L}\ \bibnamefont
  {Prejbeanu}}, \bibinfo {author} {\bibfnamefont {M}~\bibnamefont {Kerekes}},
  \bibinfo {author} {\bibfnamefont {R~C}\ \bibnamefont {Sousa}}, \bibinfo
  {author} {\bibfnamefont {H}~\bibnamefont {Sibuet}}, \bibinfo {author}
  {\bibfnamefont {O}~\bibnamefont {Redon}}, \bibinfo {author} {\bibfnamefont
  {B}~\bibnamefont {Dieny}}, \ and\ \bibinfo {author} {\bibfnamefont {J~P}\
  \bibnamefont {Nozi{\`{e}}res}},\ }\bibfield  {title} {\enquote {\bibinfo
  {title} {Thermally assisted {MRAM}},}\ }\href {\doibase
  10.1088/0953-8984/19/16/165218} {\bibfield  {journal} {\bibinfo  {journal}
  {Journal of Physics: Condensed Matter}\ }\textbf {\bibinfo {volume} {19}},\
  \bibinfo {pages} {165218} (\bibinfo {year} {2007})}\BibitemShut {NoStop}%
\bibitem [{\citenamefont {Kadigrobov}\ \emph {et~al.}(2010)\citenamefont
  {Kadigrobov}, \citenamefont {Andersson}, \citenamefont {Radi{\'{c}}},
  \citenamefont {Shekhter}, \citenamefont {Jonson},\ and\ \citenamefont
  {Korenivski}}]{Kadigrobov2010}%
  \BibitemOpen
  \bibfield  {author} {\bibinfo {author} {\bibfnamefont {A.~M.}\ \bibnamefont
  {Kadigrobov}}, \bibinfo {author} {\bibfnamefont {S.}~\bibnamefont
  {Andersson}}, \bibinfo {author} {\bibfnamefont {D.}~\bibnamefont
  {Radi{\'{c}}}}, \bibinfo {author} {\bibfnamefont {R.~I.}\ \bibnamefont
  {Shekhter}}, \bibinfo {author} {\bibfnamefont {M.}~\bibnamefont {Jonson}}, \
  and\ \bibinfo {author} {\bibfnamefont {V.}~\bibnamefont {Korenivski}},\
  }\bibfield  {title} {\enquote {\bibinfo {title} {Thermoelectrical
  manipulation of nanomagnets},}\ }\href {\doibase 10.1063/1.3437054}
  {\bibfield  {journal} {\bibinfo  {journal} {Journal of Applied Physics}\
  }\textbf {\bibinfo {volume} {107}},\ \bibinfo {pages} {123706} (\bibinfo
  {year} {2010})}\BibitemShut {NoStop}%
\bibitem [{\citenamefont {Kadigrobov}\ \emph {et~al.}(2012)\citenamefont
  {Kadigrobov}, \citenamefont {Andersson}, \citenamefont {Park}, \citenamefont
  {Radi{\'{c}}}, \citenamefont {Shekhter}, \citenamefont {Jonson},\ and\
  \citenamefont {Korenivski}}]{Kadigrobov2012}%
  \BibitemOpen
  \bibfield  {author} {\bibinfo {author} {\bibfnamefont {A.~M.}\ \bibnamefont
  {Kadigrobov}}, \bibinfo {author} {\bibfnamefont {S.}~\bibnamefont
  {Andersson}}, \bibinfo {author} {\bibfnamefont {Hee~Chul}\ \bibnamefont
  {Park}}, \bibinfo {author} {\bibfnamefont {D.}~\bibnamefont {Radi{\'{c}}}},
  \bibinfo {author} {\bibfnamefont {R.~I.}\ \bibnamefont {Shekhter}}, \bibinfo
  {author} {\bibfnamefont {M.}~\bibnamefont {Jonson}}, \ and\ \bibinfo {author}
  {\bibfnamefont {V.}~\bibnamefont {Korenivski}},\ }\bibfield  {title}
  {\enquote {\bibinfo {title} {Thermal-magnetic-electric oscillator based on
  spin-valve effect},}\ }\href {\doibase 10.1063/1.3686735} {\bibfield
  {journal} {\bibinfo  {journal} {Journal of Applied Physics}\ }\textbf
  {\bibinfo {volume} {111}},\ \bibinfo {pages} {044315} (\bibinfo {year}
  {2012})}\BibitemShut {NoStop}%
\bibitem [{\citenamefont {Polishchuk}\ \emph
  {et~al.}(2017{\natexlab{a}})\citenamefont {Polishchuk}, \citenamefont
  {Tykhonenko-Polishchuk}, \citenamefont {Kravets},\ and\ \citenamefont
  {Korenivski}}]{Polishchuk2017}%
  \BibitemOpen
  \bibfield  {author} {\bibinfo {author} {\bibfnamefont {D.~M.}\ \bibnamefont
  {Polishchuk}}, \bibinfo {author} {\bibfnamefont {Yu.~O.}\ \bibnamefont
  {Tykhonenko-Polishchuk}}, \bibinfo {author} {\bibfnamefont {A.~F.}\
  \bibnamefont {Kravets}}, \ and\ \bibinfo {author} {\bibfnamefont
  {V.}~\bibnamefont {Korenivski}},\ }\bibfield  {title} {\enquote {\bibinfo
  {title} {Thermal switching of indirect interlayer exchange in magnetic
  multilayers},}\ }\href {\doibase 10.1209/0295-5075/118/37006} {\bibfield
  {journal} {\bibinfo  {journal} {{EPL} (Europhysics Letters)}\ }\textbf
  {\bibinfo {volume} {118}},\ \bibinfo {pages} {37006} (\bibinfo {year}
  {2017}{\natexlab{a}})}\BibitemShut {NoStop}%
\bibitem [{\citenamefont {Polishchuk}\ \emph
  {et~al.}(2017{\natexlab{b}})\citenamefont {Polishchuk}, \citenamefont
  {Tykhonenko-Polishchuk}, \citenamefont {Holmgren}, \citenamefont {Kravets},\
  and\ \citenamefont {Korenivski}}]{Polishchuk2017a}%
  \BibitemOpen
  \bibfield  {author} {\bibinfo {author} {\bibfnamefont {D.~M.}\ \bibnamefont
  {Polishchuk}}, \bibinfo {author} {\bibfnamefont {Yu.~O.}\ \bibnamefont
  {Tykhonenko-Polishchuk}}, \bibinfo {author} {\bibfnamefont {E.}~\bibnamefont
  {Holmgren}}, \bibinfo {author} {\bibfnamefont {A.~F.}\ \bibnamefont
  {Kravets}}, \ and\ \bibinfo {author} {\bibfnamefont {V.}~\bibnamefont
  {Korenivski}},\ }\bibfield  {title} {\enquote {\bibinfo {title} {Thermally
  induced antiferromagnetic exchange in magnetic multilayers},}\ }\href
  {\doibase 10.1103/physrevb.96.104427} {\bibfield  {journal} {\bibinfo
  {journal} {Physical Review B}\ }\textbf {\bibinfo {volume} {96}},\ \bibinfo
  {pages} {104427} (\bibinfo {year} {2017}{\natexlab{b}})}\BibitemShut
  {NoStop}%
\bibitem [{\citenamefont {Kravets}\ \emph {et~al.}(2017)\citenamefont
  {Kravets}, \citenamefont {Polishchuk}, \citenamefont {Pashchenko},
  \citenamefont {Tovstolytkin},\ and\ \citenamefont
  {Korenivski}}]{Kravets2017}%
  \BibitemOpen
  \bibfield  {author} {\bibinfo {author} {\bibfnamefont {A.~F.}\ \bibnamefont
  {Kravets}}, \bibinfo {author} {\bibfnamefont {D.~M.}\ \bibnamefont
  {Polishchuk}}, \bibinfo {author} {\bibfnamefont {V.~A.}\ \bibnamefont
  {Pashchenko}}, \bibinfo {author} {\bibfnamefont {A.~I.}\ \bibnamefont
  {Tovstolytkin}}, \ and\ \bibinfo {author} {\bibfnamefont {V.}~\bibnamefont
  {Korenivski}},\ }\bibfield  {title} {\enquote {\bibinfo {title}
  {Current-driven thermo-magnetic switching in magnetic tunnel junctions},}\
  }\href {\doibase 10.1063/1.5009577} {\bibfield  {journal} {\bibinfo
  {journal} {Applied Physics Letters}\ }\textbf {\bibinfo {volume} {111}},\
  \bibinfo {pages} {262401} (\bibinfo {year} {2017})}\BibitemShut {NoStop}%
\bibitem [{\citenamefont {Torrejon}\ \emph {et~al.}(2017)\citenamefont
  {Torrejon}, \citenamefont {Riou}, \citenamefont {Araujo}, \citenamefont
  {Tsunegi}, \citenamefont {Khalsa}, \citenamefont {Querlioz}, \citenamefont
  {Bortolotti}, \citenamefont {Cros}, \citenamefont {Yakushiji}, \citenamefont
  {Fukushima}, \citenamefont {Kubota}, \citenamefont {Yuasa}, \citenamefont
  {Stiles},\ and\ \citenamefont {Grollier}}]{Torrejon2017}%
  \BibitemOpen
  \bibfield  {author} {\bibinfo {author} {\bibfnamefont {Jacob}\ \bibnamefont
  {Torrejon}}, \bibinfo {author} {\bibfnamefont {Mathieu}\ \bibnamefont
  {Riou}}, \bibinfo {author} {\bibfnamefont {Flavio~Abreu}\ \bibnamefont
  {Araujo}}, \bibinfo {author} {\bibfnamefont {Sumito}\ \bibnamefont
  {Tsunegi}}, \bibinfo {author} {\bibfnamefont {Guru}\ \bibnamefont {Khalsa}},
  \bibinfo {author} {\bibfnamefont {Damien}\ \bibnamefont {Querlioz}}, \bibinfo
  {author} {\bibfnamefont {Paolo}\ \bibnamefont {Bortolotti}}, \bibinfo
  {author} {\bibfnamefont {Vincent}\ \bibnamefont {Cros}}, \bibinfo {author}
  {\bibfnamefont {Kay}\ \bibnamefont {Yakushiji}}, \bibinfo {author}
  {\bibfnamefont {Akio}\ \bibnamefont {Fukushima}}, \bibinfo {author}
  {\bibfnamefont {Hitoshi}\ \bibnamefont {Kubota}}, \bibinfo {author}
  {\bibfnamefont {Shinji}\ \bibnamefont {Yuasa}}, \bibinfo {author}
  {\bibfnamefont {Mark~D.}\ \bibnamefont {Stiles}}, \ and\ \bibinfo {author}
  {\bibfnamefont {Julie}\ \bibnamefont {Grollier}},\ }\bibfield  {title}
  {\enquote {\bibinfo {title} {Neuromorphic computing with nanoscale spintronic
  oscillators},}\ }\href {\doibase 10.1038/nature23011} {\bibfield  {journal}
  {\bibinfo  {journal} {Nature}\ }\textbf {\bibinfo {volume} {547}},\ \bibinfo
  {pages} {428--431} (\bibinfo {year} {2017})}\BibitemShut {NoStop}%
\bibitem [{\citenamefont {Cheng}\ \emph
  {et~al.}(2018{\natexlab{a}})\citenamefont {Cheng}, \citenamefont {Xiao},\
  and\ \citenamefont {Zhu}}]{Cheng2018b}%
  \BibitemOpen
  \bibfield  {author} {\bibinfo {author} {\bibfnamefont {Ran}\ \bibnamefont
  {Cheng}}, \bibinfo {author} {\bibfnamefont {Di}~\bibnamefont {Xiao}}, \ and\
  \bibinfo {author} {\bibfnamefont {Jian~Gang}\ \bibnamefont {Zhu}},\
  }\bibfield  {title} {\enquote {\bibinfo {title} {{Interlayer Couplings
  Mediated by Antiferromagnetic Magnons}},}\ }\href {\doibase
  10.1103/PhysRevLett.121.207202} {\bibfield  {journal} {\bibinfo  {journal}
  {Physical Review Letters}\ }\textbf {\bibinfo {volume} {121}},\ \bibinfo
  {pages} {207202} (\bibinfo {year} {2018}{\natexlab{a}})}\BibitemShut
  {NoStop}%
\bibitem [{\citenamefont {Cheng}\ \emph
  {et~al.}(2018{\natexlab{b}})\citenamefont {Cheng}, \citenamefont {Xiao},\
  and\ \citenamefont {Zhu}}]{Cheng2018}%
  \BibitemOpen
  \bibfield  {author} {\bibinfo {author} {\bibfnamefont {Ran}\ \bibnamefont
  {Cheng}}, \bibinfo {author} {\bibfnamefont {Di}~\bibnamefont {Xiao}}, \ and\
  \bibinfo {author} {\bibfnamefont {Jian-Gang}\ \bibnamefont {Zhu}},\
  }\bibfield  {title} {\enquote {\bibinfo {title} {{Antiferromagnet-based
  magnonic spin-transfer torque}},}\ }\href {\doibase
  10.1103/PhysRevB.98.020408} {\bibfield  {journal} {\bibinfo  {journal}
  {Physical Review B}\ }\textbf {\bibinfo {volume} {98}},\ \bibinfo {pages}
  {020408} (\bibinfo {year} {2018}{\natexlab{b}})}\BibitemShut {NoStop}%
\bibitem [{Note1()}]{Note1}%
  \BibitemOpen
  \bibinfo {note} {Even though the structure is metallic, F-F* correlations
  across the AF spacer due to spin-polarized \protect \emph {conduction
  electrons} can be neglected since the corresponding spin relaxation length in
  AF is $\lesssim 1$ nm, much shorter than the AF layer thickness range used in
  this work; see Merodio \protect \textit {et al.} \protect \href
  {https://doi.org/10.1063/1.4862971}{Appl. Phys. Lett. \protect \textbf {104},
  032406 (2014)}; Bass \protect \textit {et al.} \protect \href
  {https://doi.org/10.1016/j.jmmm.2015.12.011}{J. Magn. Magn. Mater. \protect
  \textbf {408}, 244 (2016)}.}\BibitemShut {Stop}%
\bibitem [{\citenamefont {Gomonay}\ \emph {et~al.}(2018)\citenamefont
  {Gomonay}, \citenamefont {Yamamoto},\ and\ \citenamefont
  {Sinova}}]{Gomonay2018a}%
  \BibitemOpen
  \bibfield  {author} {\bibinfo {author} {\bibfnamefont {Olena}\ \bibnamefont
  {Gomonay}}, \bibinfo {author} {\bibfnamefont {Kei}\ \bibnamefont {Yamamoto}},
  \ and\ \bibinfo {author} {\bibfnamefont {Jairo}\ \bibnamefont {Sinova}},\
  }\bibfield  {title} {\enquote {\bibinfo {title} {{Spin caloric effects in
  antiferromagnets assisted by an external spin current}},}\ }\href {\doibase
  10.1088/1361-6463/aac56b} {\bibfield  {journal} {\bibinfo  {journal} {Journal
  of Physics D: Applied Physics}\ }\textbf {\bibinfo {volume} {51}},\ \bibinfo
  {pages} {264004} (\bibinfo {year} {2018})}\BibitemShut {NoStop}%
\bibitem [{\citenamefont {Offi}\ \emph {et~al.}(2002)\citenamefont {Offi},
  \citenamefont {Kuch},\ and\ \citenamefont {Kirschner}}]{Offi2002}%
  \BibitemOpen
  \bibfield  {author} {\bibinfo {author} {\bibfnamefont {F.}~\bibnamefont
  {Offi}}, \bibinfo {author} {\bibfnamefont {W.}~\bibnamefont {Kuch}}, \ and\
  \bibinfo {author} {\bibfnamefont {J.}~\bibnamefont {Kirschner}},\ }\bibfield
  {title} {\enquote {\bibinfo {title} {Structural and magnetic properties of
  {Fe$_x$Mn$_{1-x}$} thin films on {Cu}(001) and on {Co}/{Cu}(001)},}\ }\href
  {\doibase 10.1103/physrevb.66.064419} {\bibfield  {journal} {\bibinfo
  {journal} {Physical Review B}\ }\textbf {\bibinfo {volume} {66}},\ \bibinfo
  {pages} {064419} (\bibinfo {year} {2002})}\BibitemShut {NoStop}%
\bibitem [{\citenamefont {Pan}\ \emph {et~al.}(2001)\citenamefont {Pan},
  \citenamefont {Chen}, \citenamefont {Long}, \citenamefont {Tong},
  \citenamefont {Lu}, \citenamefont {Du}, \citenamefont {Hu},\ and\
  \citenamefont {Zhai}}]{Pan2001}%
  \BibitemOpen
  \bibfield  {author} {\bibinfo {author} {\bibfnamefont {M.H}\ \bibnamefont
  {Pan}}, \bibinfo {author} {\bibfnamefont {J}~\bibnamefont {Chen}}, \bibinfo
  {author} {\bibfnamefont {J.G}\ \bibnamefont {Long}}, \bibinfo {author}
  {\bibfnamefont {L.N}\ \bibnamefont {Tong}}, \bibinfo {author} {\bibfnamefont
  {M}~\bibnamefont {Lu}}, \bibinfo {author} {\bibfnamefont {J}~\bibnamefont
  {Du}}, \bibinfo {author} {\bibfnamefont {A}~\bibnamefont {Hu}}, \ and\
  \bibinfo {author} {\bibfnamefont {H.R}\ \bibnamefont {Zhai}},\ }\bibfield
  {title} {\enquote {\bibinfo {title} {A 90$\degree$ ferromagnetic layer
  coupling in {FM}/{AFM}/{FM} structures},}\ }\href {\doibase
  10.1016/s0304-8853(00)01140-9} {\bibfield  {journal} {\bibinfo  {journal}
  {Journal of Magnetism and Magnetic Materials}\ }\textbf {\bibinfo {volume}
  {226-230}},\ \bibinfo {pages} {1817--1819} (\bibinfo {year}
  {2001})}\BibitemShut {NoStop}%
\bibitem [{\citenamefont {Lenz}\ \emph {et~al.}(2007)\citenamefont {Lenz},
  \citenamefont {Zander},\ and\ \citenamefont {Kuch}}]{Lenz2007}%
  \BibitemOpen
  \bibfield  {author} {\bibinfo {author} {\bibfnamefont {K.}~\bibnamefont
  {Lenz}}, \bibinfo {author} {\bibfnamefont {S.}~\bibnamefont {Zander}}, \ and\
  \bibinfo {author} {\bibfnamefont {W.}~\bibnamefont {Kuch}},\ }\bibfield
  {title} {\enquote {\bibinfo {title} {Magnetic proximity effects in
  antiferromagnet/ferromagnet bilayers: The impact on the n{\'{e}}el
  temperature},}\ }\href {\doibase 10.1103/physrevlett.98.237201} {\bibfield
  {journal} {\bibinfo  {journal} {Physical Review Letters}\ }\textbf {\bibinfo
  {volume} {98}},\ \bibinfo {pages} {237201} (\bibinfo {year}
  {2007})}\BibitemShut {NoStop}%
\end{thebibliography}%






\end{document}